\documentclass[prb,aps]{revtex4}

\usepackage{bm}

 \usepackage{amsmath,amssymb}

 \def\S{{\mathfrak  S}}
 \def\Z{{\mathbb Z}}
 \def\N{{\mathbb N}}
\def\X{{\mathbb X}}
\def\Y{{\mathbb Y}}
\newcommand{\be}{\begin{equation}}
\newcommand{\ee}{\end{equation}}
\newcommand{\bea}{\begin{eqnarray}}
\newcommand{\eea}{\end{eqnarray}}
\newcommand{\beb}{\begin{eqnarray*}}
\newcommand{\eeb}{\end{eqnarray*}}

\begin{document}

\title{HIGHEST WEIGHT MACDONALD AND JACK POLYNOMIALS}

\author{Th.~Jolicoeur}
\affiliation{LPTMS, CNRS et Universit\'{e} Paris-Sud, 91405 Orsay, France }

\author{J. G. Luque}
\affiliation{Laboratoire LITIS - EA 4108, Universit\'{e} de Rouen, Avenue de l'Universit\'{e} - BP 8
76801 Saint-\'{E}tienne-du-Rouvray Cedex, France }

\begin{abstract}
Fractional quantum Hall states of particles in the lowest Landau levels
are described by multivariate polynomials. 
The incompressible liquid states when described on a sphere are fully invariant
under the rotation group. Excited quasiparticle/quasihole states are member
of multiplets under the rotation group and generically there is a nontrivial
highest weight member of the multiplet from which all states can be constructed.
Some of the trial states proposed in the 
literature belong to classical families of symmetric polynomials. In this
paper we study Macdonald and Jack polynomials that are highest weight states.
For Macdonald polynomials it is a (q,t)-deformation of the raising angular
momentum operator that defines the highest weight condition. By specialization
of the parameters we obtain a classification of the highest weight Jack polynomials.
Our results are valid in the case of staircase and rectangular partition indexing the
polynomials.
\end{abstract}

\maketitle

\section{Introduction}
The fractional quantum Hall effect is a state of electronic matter with elusive
physical properties. Its theoretical description pioneered by Laughlin~\cite{Laugh}
is based on explicit wavefunctions describing the full many-body state of the interacting
electrons. These wavefunctions are generically given by a polynomial function of the coordinates
of the particles in the plane. They have been studied in several geometries. In this paper
we are concerned by the case of the sphere and the unbounded plane.
In the plane a generic quantum state is given by an antisymmetric polynomial in 
the complex coordinates $z_1,\dots ,z_N$ (where $N$ is the number of particles) 
times a universal Gaussian factor which is 
independent of the state under consideration
and hence can be omitted completely. On the sphere a quantum state is an antisymmetric
polynomial in the spinor coordinates $(u_i,v_i)$ with 
$u_i = \cos (\theta_i/2) \exp (-i\phi_i/2)$ and
$v_i = \sin (\theta_i/2) \exp (i\phi_i/2)$ where
$\theta_i$ and $\phi_i$ are usual spherical coordinates. The polynomials appearing in these 
two cases can be related through stereographic projection~\cite{FOC}. 
Since one always factor out $\prod_{i<j}(z_i-z_j)$ from an antisymmetric polynomial
one can consider only symmetric polynomials which are acceptable wavefunctions for bosons.

In the physics of the fractional quantum Hall effect, the special polynomials that are 
relevant~\cite{DasSarma,Heinonen,JJBook}
are not in general solutions of the true eigenvalue problem involving the Coulomb interaction
between electrons. Instead they are thought~\cite{hal} to be adiabatically related to the true eigenstates.
In some cases they are exact eigenstates of some auxiliary operator.
The most prominent example is the celebrated Laughlin wavefunction whose
explicit formula is~:
\be
\Psi_L = \prod_{i<j}(z_i-z_j)^3 .
\ee
This quantum state is known to be an excellent approximation of the true state of electrons
when the lowest Landau level has filling factor 1/3. This polynomial is 
the eigenvector of smallest degree
with zero eigenvalue for the interaction energy 
$V({\bf r}-{\bf r^\prime})=\Delta_{\bf r}\delta^2 ({\bf r}-{\bf r^\prime})$.
Another prominent polynomial is the Moore-Read Pfaffian~\cite{mr}
 state~:
\be
\Psi_{\rm MR} = {\rm Pf}(\frac{1}{z_k-z_l})\prod_{i<j}(z_i-z_j) ,
\ee
where $\rm Pf$ stands for the Pfaffian of the matrix $1/(z_k-z_l)$. Here we have written it for 
bosons i.e. as a symmetric polynomial. This state is one of the candidates
to describe the elusive physics at filling factor 5/2. It is also the polynomial of
smallest degree that is an eigenvector with zero eigenvalue of the following operator~:
\be
{\mathcal H}^{(3)}=\sum_{i<j<k}\delta^2 ({\bf r}_i-{\bf r}_j)
\delta^2 ({\bf r}_j-{\bf r}_k).
\ee
This operator forbids three bosonic particles to be in the place. There is a natural
generalization to $p$ particles and the corresponding wavefunctions of smallest degree
are called the Read-Rezayi (RR) states~\cite{Read96,Read99}. These wavefunctions are thus
multivariate symmetric polynomials with special vanishing properties. Their study
was pioneered by Feigin, Jimbo, Miwa and Mukhin~\cite{FJMM1,FJMM2} and it was realized
that they belong to the family of Jack polynomials~\cite{BH1,BH2,BH3,BH4,BGS}.
Such polynomials depend upon a parameter and a partition of an integer~\cite{Macdo}.
A given polynomial
can be expanded in powers of the $z_i$ coordinates and a general term in the expansion
is characterized by the set of occupation numbers of the one-body orbitals
\{$n_m, m=0,1,2,\dots $\}. Note that we consider bosonic quantum Hall states
for which one can have $n_m >1$.
A given configuration of occupation numbers $(n_0,n_1,n_2\dots )$ characterizes
each term of the expansion. The set of occupation numbers defines
a partition of the integer N since $N=\sum_m n_m$
(for convenience and according to the standard notations used in literature related to symmetric 
functions, we consider {\it decreasing} partitions $\lambda=(\lambda_1,\dots,\lambda_k)$ with 
$\lambda_1\geq\dots\geq\lambda_k\geq0$).
Alternatively
one can also specify the same configuration by giving all the $m$ values that
appear with nonzero occupation numbers $(m_1..m_1m_2..m_2\dots )$
where each $m$ is repeated $n_m$ times. This set of numbers then defines
equivalently a partition of the total angular momentum $L_z=\sum_m mn_m$.
In the physics literature it is common to specify the set of occupation numbers
while the mathematical literature~\cite{Macdo} on symmetric polynomials uses instead
the partitioning of $L_z$.
 A partition $\lambda$ defines also a unique symmetric monomial
$m_\lambda$ given by~:
\begin{equation}
 m_\lambda = z_1^{k_1} \dots z_N^{k_N}+\mathrm{permutations}.
\end{equation}
This can be considered as a (unnormalized) wavefunction for N bosons in the lowest Landau level
where the quantum numbers $k_i$ of occupied orbitals can be associated
in a one to one correspondence to a set of occupation numbers \{$n_m$\}.
For example the monomial for N=3
$m=z_1^2 z_2 z_3 +{\mathit perm}$
is defined by the partition (0210$\dots$)
 since there are two bosons in the
m=1 orbital and one boson in the m=2 orbital. An arbitrary bosonic
WF in the LLL can be expanded in terms of such monomials, each of them being indexed
by a partition~:
\begin{equation}
    f =\sum_\lambda c_\lambda m_\lambda ,
\label{expans}
\end{equation}
 where $c_\lambda$ are some coefficients. For a given polynomial it may happen that not
 all partitions appear in the expansion above. Indeed there is a partial ordering on partitions
called the dominance ordering~: let $\lambda$ and $\mu$ two (decreasing) partitions then
$\lambda \geq \mu$ if $\lambda_1+\dots+\lambda_i \geq \mu_1+\dots +\mu_i$ for all $i$.
This is only a partial order~: it may happen that the relation above does not
allow comparison of two partitions. Some of the trial wavefunctions proposed in the
FQHE literature have the property that there is a dominant partition with respect to
this special order and all partitions appearing in the expansion Eq.(\ref{expans})
are dominated by a leading one~:
\begin{equation}
 \Psi = \sum_{\mu \leq \lambda} c_\mu m_\mu .
\end{equation}
This was first noted by Haldane and Rezayi~\cite{HRsqueezed} in the
case of the Laughlin wavefunction.  
The dominance plays a key role in the expansion of powers
of the VanderMonde determinant~\cite{FGIL,KTW}. This property of dominance is
also shared by many special orthogonal polynomials in several
variables~\cite{Macdo}. The Jack polynomials noted
$J^{\alpha}_{\lambda}$ are a family indexed by a partition $\lambda$
which dominates the expansion in monomials
and depend upon one parameter $\alpha$. In fact we have~:
\begin{equation}
 \Psi_{RR}^{(k)}= {\mathcal S}
\prod_{i_1<j_1}(z_{i_1}-z_{j_1})^2  \dots
\prod_{i_k<j_k}(z_{i_k}-z_{j_k})^2
\propto J^{-(k+1)}_{\lambda_k}(\{z_i\}),
\end{equation}
where the first equality defines the Read-Rezayi states, one divides the particles
into $k$ packets and  ${\mathcal S}$ means symmetrization of the product of partial Jastrow factors.
In the case of the RR states we have $\alpha = -(k+1)$ and
$\lambda_k =(k0k0k0\dots)$. The usual bosonic Laughlin wavefunction is the special subcase
when there is only one packet $k=1$ and the Moore-Read Pfaffian corresponds to the case $k=2$.
In general the filling factor of the order-$k$ RR state is $\nu =k/2$.

In the spherical geometry~\cite{FOC}, there is a natural action of the SU(2) rotation
on the quantum states. Through the stereographic correspondence it can be 
translated in an action also on the symmetric polynomials of the planar geometry.
The rotation operators are then differential operators
acting upon the particle coordinates~:
\begin{equation}
L_{+} = E_{0},
\quad L_{-} = N_{\phi}\sum_{i = 1}^{N} z_{i}  - E_{2},
\quad L_{z}
= \frac{1}{2} N N_{\phi} - E_1,
\quad  {\textrm{where}}\quad
E_{n} =  \sum_{i = 1}^{N} z_{i}^{n} \frac{\partial}{\partial
z_{i}}.
\end{equation}
Here we have introduced $N_\phi$ which is the number of flux quanta
through the sphere. The incompressible fractional quantum Hall states
are realized for a special fine tuning of $N$ and $N_\phi$.
Deviations from the ideal relation creates quasiparticle excitations on top of
the quantum Hall fluid state.
The parent incompressible fluid is invariant by rotation and hence satisfy
$L_+ \Psi = L_- \Psi = 0$~: it is a singlet under rotations.
Quasiparticle states are less symmetrical~: a one-quasiparticle state
can satisfy $L_+ \Psi = 0$ if it is located at one the poles of the sphere.
The same condition apply to states with quasiparticles or quasiholes if they
are piled up on the poles of the sphere. Mathematically these states
are highest weight in terms of representations of the rotation group, belonging
to a nontrivial multiplet.
It is thus important to find among the classical symmetric polynomials what
are those that are highest weight ones to characterize possible candidate quantum Hall states.

In this paper, we study the generalization of this problem to a (q,t)-deformation
of the raising operator $L_+$. We find two families of Macdonald polynomials
that are highest weight states. These families differ in the type of partition
that define them~: staircase and rectangular. By specialization of the parameters we find also
the Jack polynomials indexed by such partitions that are highest weight.
While we cannot fully characterize the highest weight Jacks and Macdonald polynomials,
we propose a set of conjectures.

In section \ref{MacdoJack}, we define several mathematical tools that are used in this paper and we
give general properties of Macdonald and Jack polynomials. 
The action of the raising angular momentum operator $L_+$ 
on the symmetric functions is explained in section \ref{HWSF}.
In section \ref{qdef} we introduce a two-parameter (q,t)-deformation of the angular momentum
operator $L_+$
In section \ref{HWM} we characterize the Macdonald
polynomials that satisfy the highest weight condition for rectangular and staircase
partitions. We specialize this result to Jack polynomials in section \ref{HWJ}.
Finally section \ref{conclusion} contains our conclusions.
Two appendices are devoted to the study of some eigenproblems using the method
of section \ref{HWSF}.


\section{Macdonald and Jack polynomials\label{MacdoJack}}

Let us start this paper with a brief account of Macdonald
polynomial theory. The introduction of this two parameter family
of symmetric polynomials in this context is motivated by the fact
that Jack polynomials can be considered as a
degenerate case of the Macdonald polynomials. Hence, properties
on Jack polynomials can be studied in a more general way when
stated in terms of Macdonald polynomials. 

\subsection{Symmetric functions}

The symmetric functions in $n$ independent  variables
$X=\{x_1,\dots,x_n\}$ form the subalgebra $\Lambda_n$ of the free
algebras $K[x_1,\dots,x_n]$ ($K$ being a fixed ring) composed by
polynomials which are invariant under the action of the symmetric
group $\S_n$ consisting in permuting the variables
\be
\Lambda_n=K[x_1,\dots,x_n]^{\S_n}.
\ee
The ring $\Lambda$ of symmetric functions  in countably many
independent $x_1,x_2,\dots$ is an algebra obtained by applying the
projective limit (see Ref.(\onlinecite{Macdo}) I.2 for more details)
\be
 \Lambda=\lim_{\longleftarrow}\Lambda_n.
\ee
When $K$ is a field, the space $\Lambda$ has its bases indexed by
partitions.
 For convenience  a partition will denoted by a decreasing
vectors in $\N$ (as in Refs.(\onlinecite{Macdo,Lasc})).

 When the number of variables is finite, there is an other system
 of notations (see \emph{e.g.} Refs.(\onlinecite{BH1,BH2,BH3}))
 that takes account of the interpretation of partitions as  
vectors of occupation levels of a system
 of particles.
We will use Latin letters $u,\ v,\ w,\dots$ for these notation
instead of Greek letters $\lambda,\ \mu,\ \nu,\dots$ which will be
reserved for the decreasing vectors. The ``translation'' between
the two systems of notations is very easy to understand while the
$v_i$ component of the vectors $v$ equals the number of parts
equal to $i$ in $\lambda$. For instance, for $6$ variables
(particles) one has
\be
\lambda=[4421]\sim v=[2,1,1,0,2,0,\dots]
\ee
Note also that the classical results on symmetric functions are
clearer when stated in the decreasing vector notation.

There are three main multiplicative bases for $\Lambda$~: The power
sums $p^\lambda(\X)=p_{\lambda_1}(\X)\dots p_{\lambda_n}(\X)$
where $p_k(\X)=\sum_i x_i^k$, the elementary functions
$e^\lambda(\X)=e_{\lambda_1}(\X)\dots e_{\lambda_n}(\X)$ where
$$e_k(\X)=\sum_{x_{i_1},\dots,x_{i_n}\ distincts} x_{i_1}\dots
x_{i_n}$$ and the complete functions
$h^\lambda(\X)=h_{\lambda_1}(\X)\dots h_{\lambda_n}(\X)$ where
$$h_k(\X)=\sum_{x_{i_1},\dots,x_{i_n}} x_{i_1}\dots x_{i_n}.$$
There is also non multiplicative bases such as monomial functions
$m_\lambda(\X)$ and Schur functions $s_\lambda(\X)$ (see {\it
e.g.} Refs.(\onlinecite{Macdo,Lasc})).
 When there is no relation between the
variables (this implies that the number of variables is infinite),
the algebra of symmetric functions is a polynomial algebra over
the elementary functions and also over the homogeneous functions.

\subsection{The $\lambda$-ring notations}

The generating function of the complete symmetric functions for an
alphabet $\X$ has a well-known factorized expression
\be
\sigma_z(\X):=\sum_{i}h_i(\X)z^i=\prod_{x\in\X}\frac1{1-xz}.
\ee
If $\Y$ is an alphabet disjoint of $\X$, straightforwardly
\be
\sigma_z(\X\cup\Y)=\sigma_z(\X)\sigma_z(\Y).
\ee
But more generally, if $\Y$ contains some letters of $\X$,
\begin{equation}
\label{unionXY}
\sigma_z(\X)\sigma_z(\Y)=\sigma_z(\X\cup\Y)\sigma_z(\X\cap\Y).
\end{equation}
In fact it is more convenient to consider an alphabet, not as a
set of variables, but as the formal sum of its variables,
$\X=x_1+x_2+\dots$.
In this case, Eq. (\ref{unionXY}) reads~:
\begin{equation}
\label{sumXY}
\sigma_z(\X)\sigma_z(\Y)=\sigma_z(\X+\Y).
\end{equation}
Hence, the binary operator $+$ acting on the alphabets will encode
the transformation sending $h_i(\X)$ to $\sum_{j+k=i}
h_j(\X)h_k(\Y)$. More generally, the product of an alphabet $\X$
by an element $\alpha$ of the ground field allows to define a new
alphabet denoted by $\alpha\X$ and whose complete functions are
given by~:
\be
 \sigma_z(\alpha\X)=\sigma_z(\X)^\alpha.
\ee
For example, if $\alpha=-1$, one has~:
\be
 \sigma_z(-\X)=\sigma_z(\X)^{-1}=\prod_z(1-xz)=
\sum_i(-1)^ie_i(\X)z^i=\lambda_{-z}(\X),
\ee
where $\lambda_z(\X)=\sum_i e_i(\X)z^i$ is the generating function
of the elementary functions. The operation sending $\X$ on $-\X$
can be formally interpreted as the transformation sending
$h_i(\X)$ on $\pm e_i(\X)$.\\
The last operation we introduce is the multiplication of two
alphabets $\X\Y$. The transformation is clearer when stated in
terms of power sums~:
\be
 p_\lambda(\X\Y)=p_\lambda(\X)p_\lambda(\Y).
\ee
For interpreting this operation on the complete function, we need
to introduce the generating function of the power sums
\be
 \Psi_z(\X):=\sum_i {p_i(\X)\over i}z^i=\log\sigma_z(\X).
\ee
After a short computation, one finds~:
\be
 \sigma_z(\X\Y)=
\sum_i\sum_{\lambda\vdash i}z_\lambda^{-1}p^\lambda(\X)p^\lambda(\Y)z^i
\ee
where $\lambda\vdash i$ means that $\lambda$ is a partition of
weight $i$ and $z_\lambda=\prod_{i\geq 1}i^{m_i}m_i!$ if $m_i$
denotes the number of parts of $\lambda$ equal to $i$.
The function $K(\X,\Y)=\sigma_1(\X\Y)$ has another important role
in the theory of symmetric functions~: it is the reproducing kernel
of the usual scalar product whose evaluation on power sums is~:
\begin{equation}
\label{usual}
 \langle p_\lambda,p_\mu\rangle=z_\lambda\delta_{\mu,\lambda}.
\end{equation}
By reproducing kernel, we mean that if ($B_\lambda)_\lambda$ and
$(C_\lambda)_\lambda$ are two bases in duality for the scalar
product $\langle\,,\,\rangle$ ( $\langle
B_\lambda,C_\mu\rangle=\delta_{\mu,\lambda}$) then we have~:
\be
 \langle K(\X,\Y),B_\lambda(X)\rangle_X = B_\lambda(\Y).
\ee
The Schur basis is the unique basis $(s_\lambda)$ orthonormal for
$\langle\,,\,\rangle$ verifying that the dominant monomial in
$s_\lambda$ is $x_1^{\lambda_1}\dots x_n^{\lambda_n}$, hence~:
\be
 K(\X,\Y)=\sum_\lambda s_\lambda(\X)s_\lambda(\Y).
\ee

\subsection{Macdonald Polynomials}

The Macdonald polynomials $(P_\lambda(\X;q,t))_\lambda$ form the
unique basis of symmetric functions orthogonal for the standard
$(q,t)$-deformation of the usual scalar product on symmetric
functions~:
\begin{equation}
\label{qtscalar}
\langle p^\lambda,p^\mu\rangle=\prod_{i=1}^{l(\lambda)}
{1-q^{\lambda_i}\over
1-t^{\lambda_i}}z_\lambda\delta_{\lambda,\mu},
\end{equation}
see \emph{e.g.} Ref. (\onlinecite{Macdo}) VI.4 p322, verifying the following equation~:
\begin{equation}
\label{PM}
P_\lambda(\X;q,t)=m_\lambda(\X)+\sum_{\mu\leq\lambda}u_{\lambda\mu}m_\mu(\X),
\end{equation}
where $m_\lambda$ is a monomial function in the notation of
Ref.(\onlinecite{Macdo}), I.2.1, p8. Their generating function is (see {\it
e.g.} Ref.(\onlinecite{Macdo}), VI.4.13, p324)~:
{\footnotesize
\begin{equation}
\label{Kqt}
K_{q,t}(\X,\Y):=\sum_\lambda\prod_{i=1}^{l(\lambda)}{1-t^{\lambda_i}\over
1-q^{\lambda_i}}p^\lambda(\X)p^\lambda(Y)=\sigma_1\left({1-t\over
1-q}\X\Y\right)=\sum_\lambda
P_{\lambda}(\X;q,t)Q_{\lambda}(\Y;q,t),
\end{equation}
}
where~:
\begin{equation}
\label{QM}
Q_\lambda(\X;q,t)= b_\lambda(q,t)
P_\lambda(\X;q,t),
\end{equation}
with~:
\be
b_{\lambda}(q,t)=\prod_{(i,j)\in\lambda}
{1-q^{\lambda_i-j}t^{\lambda'_j-i+1}\over
1-q^{\lambda_i-j+1}t^{\lambda'_j-i}}.
\ee
Note that the first part of equality (\ref{Kqt}) is obtained by a
straightforward computation from the expression of $K_{qt}(\X,\Y)$
as the reproducing kernel of the scalar product
$\langle\,,\,\rangle$ ~:
\be
K_{qt}(\X,\Y)=\sum_\lambda {1-t^{\lambda_i}\over
1-q^{\lambda_i}}z_\lambda p^\lambda(\X)p^\lambda(\Y).
\ee
Whilst the second part is a non trivial consequence of the Pieri
formula,
see \emph{e.g.} Ref.(\onlinecite{Macdo}), VI.6.19, p339.
The notation ${1-t\over 1-q}\X\Y$ must be understood in terms of
$\lambda$-ring as a product of the three alphabets ${1-t\over
1-q},\,\X$ and $\Y$. The resulting alphabet is described by the
following formal sum~:
\be
{1-t\over 1-q}\X\Y=\sum_{x\in\X,\,y\in\Y\atop k\in\N} q^k xy -
\sum_{x\in\X,\,y\in\Y\atop k\in\N} tq^k xy.
\ee
Macdonald polynomials appears in literature with numerous
normalizations, let us recall the main ones.
 The normalization $J$ is particularly interesting since the
 coefficients of the monomials are polynomials in $q$ and $t$. It
 follows that a polynomial $J_\lambda$ has no pole (in contrast to
 the  normalization $P$). These polynomials are defined by~:
 \begin{equation}
\label{JM}
 J_\lambda(\X;q,t)=c_\lambda(q,t)
 P_\lambda(\X,q,t)=c'_\lambda(q,t)Q_\lambda(\X;q,t),
 \end{equation}
where $c_\lambda(q,t)=\prod_{(i,j)\in \lambda}
(1-q^{\lambda_i-j}t^{\lambda_j'-i+1})$ and
$c'_\lambda(q,t)=\prod_{(i,j)\in \lambda}
(1-q^{\lambda_i-j+1}t^{\lambda_j'-i})$ if $\lambda'$ denotes the
partition conjugate with $\lambda$.
As in Ref.(\onlinecite{Lass}), we will denote by $J^\sharp$ the adjoint
normalization of $J$ \emph{w.r.t} the scalar product
$\langle\,,\,\rangle$~:
\begin{equation}
\label{Jsharp}
 \langle J^\sharp_\lambda,J_\mu\rangle_{q,t}=\delta_{\lambda,\mu}.
\end{equation}
When the alphabet is finite, Lassalle  introduced~\cite{Lass} 
another normalization denoted by $J^\star$ which is defined by~:
\begin{equation}
\label{Jstar}
 J_\lambda^\star(x_1+\dots+x_n;q,t)=
{J_\lambda(x_1+\dots+x_n;q,t)\over J_\lambda({1-t^n\over 1-t};q,t)}.
\end{equation}

\subsection{Skew functions}

We describe first the general process allowing to define skew
functions from any bases of symmetric functions. Consider two
bases $(B_\lambda)_\lambda$ and $(C_\lambda)_\lambda$ which are
adjoint \emph{w.r.t.} a certain scalar product denoted by
$\{\,,\,\}$. We will denote by $K^{\{\ \}}(\X,\Y)$ the reproducing
kernel of $\{\,,\,\}$~:
\be
\,K^{\{\
\}}(\X,\Y)=\sum_{\lambda}B_\lambda(\X)C_\lambda(\Y).
\ee
The aim of the construction of skew functions deals with the
problem of the description of the operator which is adjoint of the
multiplication by $C_\lambda$,\[ \emph{i.e.}\,\{C_\lambda
P,Q\}=\{P,???.Q\},\] or equivalently to find the polynomials
$B_{\lambda/\mu}$ verifying~:
\begin{equation}
\label{defskew}
C_\mu(\Y)K^{\{\
\}}(\X,\Y)=\sum_{\lambda}B_{\lambda/\mu}(\X)C_\lambda(\Y).
\end{equation}
A straightforward computation gives the equality~:
\begin{equation}
\label{skew}
B_{\lambda/\mu}=\sum_{\mu}c^\lambda_{\mu,\nu}B_\nu ,
\end{equation}
where the $c^\lambda_{\mu,\nu}$ are the structure coefficients of
the basis $(C_\lambda)_\lambda$~:
\be 
C_\mu C_\nu=\sum_\lambda c^\lambda_{\mu,\nu}
C_\lambda.
\ee

{\bf Remark}
{\it
\label{twoscalar}
If $\{\,,\,\}'$ is a second scalar product and
$(B'_\lambda)_\lambda$ the basis adjoint to $(C_\lambda)_\lambda$
\emph{w.r.t.} $\{\,,\,\}'$, then the decomposition in the basis
$B'$ of the polynomials $B'_{\lambda/\mu}$ involves the same
coefficients than which appear in the decomposition of
$B_{\lambda/\mu}$ in the basis $B$, that is
\be
B'_{\lambda/\mu}=\sum_{\mu}c^\lambda_{\mu,\nu}B'_\nu.
\ee
}

Now, suppose that the reproducing kernel of $\{\,,\,\}$ is
multiplicative for the addition of alphabets~:
\be
 K^{\{\,\}}(\X+\Y,\Z)=K^{\{\,\}}(\X,\Z)K^{\{\,\}}(\Y,\Z).
\ee
One has then~:
\[
\begin{array}{rcl}
 K^{\{\,\}}((\X+\Y),\Z)&=&K^{\{\,\}}(\X,\Z)K^{\{\,\}}(\Y,\Z)\\
 &=&\displaystyle\sum_{\mu,\nu} B_\mu(\X)B_\nu(\Y) C_\mu(\X)C_\nu(\Y)\\
 &=&\displaystyle \sum_\lambda\left(\sum_\mu
 c^\lambda_{\mu,\nu}B_\mu(\X)B_\nu(\Y)\right)C_\lambda(\Z).
\end{array}
\]
Hence,
\begin{equation}
\label{sumskew}
\begin{array}{rcl}
B_\lambda(\X+\Y)&=&\{K^{\{\,\}}((\X+\Y)\Z),B_\lambda(\Z)\}_Z\\
&=& \displaystyle\sum_\mu B_{\lambda/\mu}(\X)B_\mu(\Y).
\end{array}
\end{equation}
In particular, this is the case for skew Schur functions
$s_{\lambda/\mu}$, skew Jack polynomials and skew Macdonald
polynomials.

\subsection{Jack polynomials}

If we set $q=t^\alpha$ and tends $t$ to $1$ in the previous
equalities one recovers the theory of Jack polynomials. Indeed,
Jack polynomials are homogeneous symmetric functions orthogonal
\emph{w.r.t.} the scalar product defined on power sums by~:
\be
 \langle p^\lambda,p^\mu\rangle_\alpha=\lim_{t\rightarrow 1} \langle
 p^\lambda,p^\mu\rangle_{t^\alpha,t}=\alpha^{l(\lambda)}z_\lambda\delta_{\lambda,\mu}.
\ee
The reproducing kernel associated to this scalar product is~:
\be
K_\alpha(\X,\Y)=\lim_{t\rightarrow1}
K_{t^\alpha,t}(\X,\Y)=\sigma_1(\alpha^{-1}\X\Y)=(\sigma_1(\X\Y))^{1\over\alpha}.
\ee
Jack polynomials appear in literature with normalization
$P_\lambda^{(\alpha)}$ (resp. $Q_\lambda^{(\alpha)}$,
$J^{(\alpha)}_\lambda$, ${J^\sharp}_\lambda^{(\alpha)}$,
${J^\star}_\lambda^{(\alpha)}$) which is deduced from $P_\lambda$
(resp. $Q_\lambda$, $J_\lambda$, ${J^\sharp}_\lambda$,
${J^\star}_\lambda$) by putting $q=t^\alpha$ and sending $t$ to
$1$ in Eq. (\ref{PM}) (resp.  Eq. (\ref{QM}), Eq. (\ref{JM}), Eq.
(\ref{Jstar}), Eq (\ref{Jsharp})). 
For more details see \emph{e.g.} Ref.(\onlinecite{Macdo}), VI. 10.
The classical case of Schur functions $s_\lambda$ is recovered
when setting $\alpha=1$ in the previous equalities.


\section{Highest weight symmetric functions\label{HWSF}}

In this paragraph, we explain briefly the algorithm of A.
Lascoux~\cite{Lasc2} to compute the action of the operator~:
\be
\label{Lplus}
        L_+=\sum_{i=1}^n{\frac{\partial} {\partial x_i}}
\ee
on symmetric functions. Its action on power sums is easily
described in terms of power sums, since we have~:
\be
 L_+.\Psi_z(\X)=nz+z^2\frac{\partial }{\partial z}\Psi_z(\X),
\ee
or equivalently $L_+.p_k(\X)=kp_{k-1}(\X)$. Since, $L_+$ is a
first order differential operator, this equality is sufficient to
describe its action on symmetric functions.
Its action on elementary symmetric functions is also very simple
to understand~:
 \be
  L_+.e_k(\X)=(n-k+1)e_{k-1}(\X),
 \ee
and is obtained from the action of $L_+$ on the generating
function $\lambda_z(\X)$~:
\be
 L_+.\lambda_z(\X)=z\sum_i\lambda_z(\X-x_i)
\ee
Since the set of elementary functions
$(e^\lambda(\X))_{\lambda,l(\lambda)\leq n}$ is a basis of the
space of symmetric functions and that these functions are
algebraically independent, the operator $L_+$ can be rewritten by
means of the operators $\partial_{e_k}:={\frac{\partial}{\partial
e_k}}$~:
\be
 L_+ = \sum_{i=1}^n (n-i+1) e_{i-1} \partial_{e_i}.
\ee
This expression is unsatisfactory because it is somewhat difficult
to cope with the coefficient $(n-i+1)$.
To simplify the problem, one introduces a new alphabet $\tilde \X$
of size $n$ which consists of the roots of the polynomial~:
  \begin{equation}
\label{tildeX}
 \tilde \lambda_z(\X)=\sum_{i=1}^n{\frac{(n-i)!}
 {n!}}e_i(\X)z^i=\lambda_z(\tilde \X).
 \end{equation}
 Note that a function is symmetric in $\X$ if and only if it is
 symmetric in $\tilde \X$. To convert an expression in $\X$ to an
 expression in $\tilde \X$, it suffices to apply the formal
 substitution $e_i(\X)={\frac{n!}{ (n-i)!}}e_i(\tilde \X)$.\\
 Setting $\tilde e_i(\X)=e_i(\tilde \X)$, the operator $L_+$ has
 the following nice expression~:
\be
 L_+ = \sum_{i=1}^n  {\tilde e_{i-1}} \partial_{\tilde e_i}.
\ee
We will use the Macdonald notation~\cite{Macdo} to denote the
basis $(c_\lambda)_\lambda$ which is the adjoint basis to the
power sum basis $(p^\lambda)_\lambda$  for the usual scalar
product (\ref{usual}). Equivalently, $c_\lambda=z^{-1}_\lambda
p^\lambda$. For simplicity, we set $\tilde
c_\lambda(\X)=c_\lambda(\tilde\X)$. With this notation, the kernel
of $L_+$ is characterized by a theorem of Mac Mahon~\cite{MacM}
corrected by Sylvester~\cite{Sylv}.

\bigskip
{\bf Proposition}. (Mac Mahon-Sylvester)

{\it
 A symmetric polynomial belongs to the kernel of $L_+$ if and only
 if its expansion in the basis $(\tilde c_\lambda(\X))_\lambda$
 does not contains any partition having a part equal to $1$.\\
 Equivalently, the kernel of $L_+$ is the subring generated by
 $\tilde c_2(\X)$, $\tilde c_3(\X),\dots$.
}
\bigskip

Note that, in the terminology of Mac Mahon, a polynomial is said
{\it semi-invariant} if and only if it belongs to the kernel of
$L_+$ while Bernevig and Haldane~\cite{BH1,BH2,BH3,BH4} use this
word for polynomials which are in the kernel of both  $L_+$ and~:
\be
    L_-=N_\phi\sum_i x_i-\sum_ix_i^2{\frac{\partial}{\partial x_i}}.
\ee
To avoid confusion, we will say that a polynomial is {\it
highest weight} (HW) if and only if it is annulled by $L_+$.

In conclusion, we want to emphasize  the relevance of the functions $ \tilde c_\lambda $  
to analyse eigenvalue problems in the context of the fractional quantum Hall effect.
In appendix I, we give a closed formula for the 
expansion of the ``yrast'' eigenfunctions of the delta function interaction
when the angular momentum equals the number of particles
in terms of $\tilde c_\lambda$.
In appendix II, 
we discuss some spectral properties of the Read-Rezayi operator~\cite{Read96,Read99}~:
\be
 {\cal
 H}^{(k)}:=\sum_{i_1<\dots<i_{k+1}}\delta^{(2)}(x_{i_1}-x_{i_1})
\dots
\delta^{(2)}(x_{i_k}-x_{i_{k+1}}).
 \ee

\section{A (q,t)-deformation of $L_+$ \label{qdef}}

In Ref.(\onlinecite{Lass}), Lassalle introduced generalized binomial
        coefficients $\left(\lambda\atop \mu\right)_{q,t}$ in the
        aim to understand the action of a $(q,t)$-deformation of
        $L_+$ on Macdonald polynomials.
        These binomial coefficients are the coefficients  of
        $J^\sharp_\lambda$ in the generating series
        \begin{equation}
\label{defbin}
        J^\sharp_\mu(\X;q,t)K_{q,t}(1+t+\dots+t^{n-1},\X)=\sum_\lambda
        t^{|\lambda|-|\mu|}\left(\lambda\atop\mu\right)_{q,t}J^\sharp_\lambda(\X;q,t).
        \end{equation}
        These coefficients are equal, up to a multiplicative
        coefficient which is a power of $t$, to skew Macdonald
        polynomials specialized to the alphabet $T_n=1+t+\dots+t^{n-1}$.
        More precisely~:
        \be
        \left(\lambda\atop\mu\right)_{q,t}=t^{|\mu|-|\lambda|}J_{\lambda/\mu}
(T_n;q,t)=\sum_{\nu}{j^\sharp}^\lambda_{\mu,\nu}(q,t)J_{\mu}(T_n;q,t)
        \ee
        where coefficients ${j^\sharp}^\lambda_{\mu,\nu}(q,t)$ denotes the
        structure coefficients of the basis
        $(J^\sharp_\lambda)_\lambda$,
        \be
        J^{\sharp}_\mu J^\sharp_\nu =
\sum_\lambda{j^\sharp}^\lambda_{\mu,\nu}J^\sharp_\lambda.\ee
        Indeed, it suffices to remark that Eq. (\ref{defbin}) is a
  special case of Eq.(\ref{defskew}), hence the result is
  obtained by applying Eq.(\ref{skew}).
The $(q,t)$-deformation of $L_+$ considered by
Lassalle~\cite{Lass} is the same as introduced by Macdonald~\cite{Macdo}
VI.3. Let us recall it here.
First, one has to define the $q$-deformation of the derivation
$\partial / \partial x_i$ by means of the divided difference~:
\be
 {\partial /\partial_q x_i} f(\X)=\frac{f(\X)-f(\X-(1-q)x_i)}{
 x_i-qx_i}.
\ee
Remember that in our notation the alphabet $\X-(1-q)x_i$ is
obtained from $\X$ by substituting $qx_i$ to $x_i$ and remark that
if $q$ tends to $1$, then ${\partial / \partial_q x_i}$ tends
to ${\partial /
\partial x_i}$. The $(q,t)$-deformation of the operator
$L_+^{q,t}$ is defined by
\begin{equation}
\label{defLqt}
 L_+^{q,t}:=\sum_{i=1}^n\prod_{j=1\atop i\neq j}^n{\frac{tx_i-x_j}
 {x_i-x_j}}{\frac{\partial}{\partial_qx_i}}.
\end{equation}
The operator $L_+$ can be recovered from Eq.(\ref{defLqt}) by
taking the limit~:
\be
 L_+=\lim_{(q,t)\rightarrow (1,1)} L_+^{q,t}.
\ee
 In Ref.(\onlinecite{Lass}), Lassalle proved the following very interesting
 identity which describes the action of $L_+^{q,t}$ on the
 Macdonald polynomials $J ^\star_\lambda$ by means of generalized
 binomial coefficients.
  For each partition $\lambda$ of length $l(\lambda)\leq n$, one
  has~:
\begin{equation}
\label{EqLqtonJstar}
  L_+^{q,t}
  J^\star_\lambda(\X;q,t)=\sum_{i}
{\lambda\choose\lambda_{(i)}}_{q,t}
J^\star_{\lambda_{(i)}}(\X;q,t),
\end{equation}
  where $\lambda_{(i)}$ denotes the
  vector obtained by subtracting $1$ to the part $i$ of the
  partition $\lambda$ and the sum runs over the integers $i$ such
  that $\lambda_{(i)}$ is a decreasing partition.
One way to understand why the coefficients
${\lambda\choose\lambda_{(i)}}_{q,t}$ appear is to introduce a
new scalar product $\{\,,\,\}$ for which $J^\star_\lambda(\X;q,t)$
and $t^{|\lambda|}J^\sharp_\lambda(\X;q,t)$ are adjoint. The
reproducing kernel of $\{\,,\,\}$ is by definition a generalized
hypergeometric function associated to the  Macdonald polynomials~:
\be
    K^{\{\,,\,\}}(\X,\Y)=\,_0{\cal F}_0(\X,\Y;q,t):=\sum_\lambda
    t^{|\lambda|}J^\sharp_\lambda(\X;q,t)J^\star_\lambda(\Y;q,t),
\ee
$\X$ and $\Y$ being two alphabets with at most $n$ letters. The
difficult part of Lassalle's reasoning consists in proving that
the operator ${L_+}^{q,t}$ and $e_1.$ are adjoint~:
\be
\emph{i.e.}\, {L_+^{q,t}}_\X.\,_0{\cal
F}_0(\X,\Y;q,t)=e_1(\Y)\,_0{\cal F}_0(\X,\Y;q,t).
\ee
This is a rather technical computation that we do not repeat here.
But, this fact being established, Remark (\ref{twoscalar}), combined
with  $e_1(\X)=J^\star_1(\X;q,t)=J^\sharp_1(\X;q,t)$, explains
completely why generalized binomial coefficients appear in 
Eq.(\ref{EqLqtonJstar}).
 In the sequel, we will use the normalization $P$ instead of
 $J^\star$. The action of $L_+^{q,t}$ on $P_\lambda(\X;q,t)$ can
 be easily deduced from Eq.(\ref{EqLqtonJstar}).
First, one plugs successively the definition of $J^\star$ 
(Eq.(\ref{Jstar})) and $J$ (Eq.(\ref{JM})) in 
Eq.(\ref{EqLqtonJstar}) and obtains~:
\be
L_+^{q,t}.P_\lambda(\X;q,t)=\sum_i{\lambda\choose\lambda_{(i)}}_{q,t}{J_\lambda(T_n;q,t)/
J_{\lambda_{(i)}}(T_n;q,t)}{c_{\lambda_{(i)}}(q,t)\over
c_{\lambda}(q,t)}P_{\lambda_{(k)}}(\X;q,t).
\ee
 Knowing the value of $J_\mu(T_n;q,t)$~:
 \be
J_\mu(T_n;q,t)=\prod_{(i,j)\in\lambda}(t^{i-1}-q^{j-1}t^n),
 \ee
 see \emph{e.g.} Ref.(\onlinecite{Macdo}), VI.8, Eq. (8), and using
 the value of $c_\lambda$ given in Eq.(\ref{JM}), one finds~:
{\footnotesize 
\begin{equation}
\label{L+qtP}
L_+^{q,t}.P_\lambda(\X;q,t)=\sum_i{1-t^{n-i}q^{\lambda_i}\over
 1-q}\prod_{j=i+1}^n{1-q^{\lambda_i-\lambda_j-1}t^{j-i+1}\over 1-q^{\lambda_i-\lambda_j}t^{j-i}}
 {1-q^{\lambda_i-\lambda_j}t^{j-i-1}\over
 1-q^{\lambda_i-\lambda_j-1}t^{j-i}}P_{\lambda_{(i)}}(\X;q,t).
\end{equation}
}
 For simplicity, a polynomial belonging in the kernel of
 $L_+^{q,t}$ will be called {\it highest weight}.

\section{Highest weight Macdonald polynomials \label{HWM}}

In this section, one investigates two families of highest
weight Macdonald polynomials and we will suppose that  $t^{k-1}q^{r+1}=1$ for some integers $k,r\in\N$.

\subsection{Weakly admissible partitions}

Let us recall some results contained in the paper of Feigin {\it
et al.}~\cite{FJMM2}. The aim of Ref.(\onlinecite{FJMM2}) is to study ideals
of polynomials defined by certain vanishing conditions (called
wheel conditions). For their purpose, Feigin \emph{et al.} defined
the notion of admissible partitions.
A partition $\lambda$ is said $(r,k,n)$-{\it admissible} if for
each $i=1\dots n-k$, one has $\lambda_i-\lambda_{i+r}\geq k$. In
this definition, one considers that the partition $\lambda$
encodes an element of a basis of the symmetric functions algebra
for an alphabet of size $n$, hence the partition $\lambda$ is
completed with $0$ at the right by setting $\lambda_i=0$ if
$i>l(\lambda)$.
They proved the following property~:
Suppose $1\leq i<j\leq n$ and $\lambda$ is a $(r,k,n)$-admissible
partition. Then one has~:
\begin{equation}
\label{FJMM1}
 q^{\lambda_i-\lambda_j}t^{j-i}\neq 1,\,  q^{\lambda_i-\lambda_j-1}t^{j-i+1}\neq
 1,\, q^{\lambda_i-\lambda_j-1}t^{j-i}\neq 1.
\end{equation}
In addition, if $\lambda_j<\lambda_{j+1}$ then
$q^{\lambda_i-\lambda_j}t^{j-i-1}\neq 1$.
This can be straightforwardly adapted to slightly more general
partitions. A partition $\lambda$ will be called {\it weakly}
$(r,k,n)$-{\it admissible} if for each $i=1,\dots, n-k$, one has
$\lambda_i-\lambda_{i+r}\geq k$ or $\lambda_i=0$.
Let $\lambda$ be weakly $(r,k,n)$-admissible partition  and $1\leq
i<j\leq l(\lambda)+r $. Then, remarking that $\lambda$ is a
$(r,k,l(\lambda)+r)$-admissible
 partition and applying Eq.(\ref{FJMM1}), one obtains~:
\begin{equation}
\label{WFJMM1}
 q^{\lambda_i-\lambda_j}t^{j-i}\neq 1,\,  q^{\lambda_i-\lambda_j-1}t^{j-i+1}\neq
 1,\, q^{\lambda_i-\lambda_j-1}t^{j-i}\neq 1.
\end{equation}
In addition, if $\lambda_j<\lambda_{j+1}$ then
$q^{\lambda_i-\lambda_j}t^{j-i-1}\neq 1$.
Feigin \emph{et al}~\cite{FJMM2} proved an interesting condition
for the poles of $P_\lambda(\X;q,t)$ when $\lambda$ is
$(r,k,n)$-admissible.

\bigskip

{\bf Lemma 1.} ({\it Feigin, Jimbo, Miwa and Mukhin})

{\it
Assume either $\lambda$ is  $(r,k,n)$-admissible or else $\lambda$
is obtained from a $(r,k,n)$-admissible partition by adding or
removing one node. Then $P_\lambda(\X;q,t)$ has no pole at
$(t,q)=\left(u^{k-1\over m},\omega_1u^{-{r+1\over m}}\right)$
where $\omega_1=\exp\left\{2i\pi(1+dm)\over r-1\right\}$ with
$m=\gcd(r+1,k-1)$ and $d\in\Z$.
}

They obtained this result by investigating the coefficients of
$P_\lambda(\X;q,t)$ in the expansion in the monomial basis. This
expansion is known (see {\it e.g.} Ref.(\onlinecite{Macdo}), VI 7, (7.13')~:
\be
 P_\lambda(\X;q,t)=\sum_T \psi_T(q,t)x^T,
\ee
where the sum runs over
tableaux $T$ of shape $\lambda$, and involves rational
fractions $\psi_T(q,t)$ which are an explicit product of quotients
of fractions given by~:
\be
b_\lambda(i,j;q,t)=
{1-q^{\lambda_i-j}t^{\lambda'_j-i+1}\over
1-q^{\lambda_i-j+1}t^{\lambda'_j-i}},
\ee
where $(i,j)$ is a node of $\lambda$. Hence, Lemma 1
remains true for weakly admissible partitions.

\bigskip

{\bf Lemma 2.}  

{\it
Assume either $\lambda$ is  weakly $(r,k,n)$-admissible or else
$\lambda$ is obtained from a $(r,k,n)$-admissible partition by
adding or removing one node. Then $P_\lambda(\X;q,t)$ has no pole
at $(t,q)=\left(u^{k-1\over m},\omega_1u^{-{r+1\over m}}\right)$.
}

{\bf Proof} The expansion of $P_\lambda(\X;q,t)$ on monomial
functions involves coefficients whose denominators are constituted
by products of $1-q^{\lambda_i-j}t^{\lambda'_j-i+1}$ or
$1-q^{\lambda_i-j+1}t^{\lambda'_j-i}$ with $(i,j)\in\lambda$. This
 implies that $i\leq l(\lambda)\leq l(\lambda)+r$. Since $\lambda$
is a $(r,k,l(\lambda)+r)$-admissible partition, the result is a
direct consequence of Lemma 1.$\Box$

Let us give an example. Consider the partition $\lambda=(2,2)$
which is weakly $(2,2,n)$-admissible for any $n>2$. The polynomial
$P_{22}(\X;q,t)$ admits the following decomposition over the
monomial functions 
{\footnotesize\[
P_{22}(\X;q,t)=m_{{2,2}}+{\frac {\left (q+1\right )\left
(-1+t\right )m_{{2,1,1}}}{qt -1}}+{\frac {\left (q+1\right )\left
(-1+t\right )^{2}\left (t+2\,qt+2 +q\right )m_{{1,1,1,1}}}{\left
(qt-1\right )\left (q{t}^{2}-1\right )}}
\]}
This equality is independent of the size $n$ of the alphabet.
Hence, the only possible poles are such that $qt=1$ or $qt^2=1$.
Lemma 2 predicts that $(t,q)=(u,-u)$ is not a pole of
$P_{22}(\X;q,t)$.

\subsection{Rectangular partitions}

    In this subsection, one investigates a family of highest
    weight Macdonald polynomials indexed by rectangular
    partition.
    More precisely, one proves the following result~:

{\bf Theorem I.}
{\it
    If $n\geq 2r$ then the Macdonald polynomial
    $P_{k^r}(x_1+\dots+x_n;q,t)$ belongs to the kernel of $L_+^{q,t}$
    for the specialization $(t,q)=(u^{k-1\over g},u^{r-1-n\over
    g}\omega_1)$ where $g=\gcd(k-1,n-r+1)$ and
    $\omega_1=\exp\left\{2i\pi(1+dg)\over k-1\right\}$ with
    $d\in\Z$.
}

\bigskip

    {\bf Proof} We start with the Lassalle identity for the
    normalization $P$ (Eq.(\ref{L+qtP})). This identity
    involves a unique Macdonald polynomial in its right hand
    side~:
    \begin{equation}
  L_+^{q,t}P_{k^r}(x_1+\dots+x_N;q,t)={1-q^k\over
1-q}{1-q^{k-1}t^{N+1-r}\over
1-q^{k-1}t}P_{k^{r-1}k-1}(x_1+\dots+x_N;q,t).
    \end{equation}
    Suppose now that $(t,q)=(u^{k-1\over s},u^{r-1-n\over
    g}\omega_1)$. First remark that the partition $(k^r)$ is weakly $(k,
    n-r,n)$-admissible when $n\geq 2r$. Hence, from Lemma
    1, the polynomial $P_{k^r}(x_1+\dots+x_N;q,t)$ is
    well defined. It follows that a necessary condition for
    $L_+^{q,t}P_{k^r}(x_1+\dots+x_N;q,t)=0$ is ${1-q^k\over
1-q}{1-q^{k-1}t^{N+1-r}\over 1-q^{k-1}t}=0$. If $n\geq 2r$, the
polynomial $1-q^{k-1}t^{N+1-r}$ is not divisible by $1-q^{k-1}t$
and vanishes for our specialization. It remains to prove that
$P_{k^{r-1}k-1}(x_1+\dots+x_N;q,t)$ has no pole at
$(t,q)=(u^{k-1\over s},u^{r-1-n\over
    g}\omega_1)$. Since the partition $(k^{r-1},k-1)$ is obtained
    from the weakly admissible  partition $(k^r)$ by subtracting
    $1$ to the last part, this is again a consequence of Lemma
    1. This ends the proof.$\Box$

    Let us give some examples to illustrate this result.
     The following polynomials are highest weight Macdonald
     polynomials,
\begin{enumerate}
 \item $P_4(x_1+x_2+x_3;\exp(2i\pi/3)u^{-1},u),$
 \item $P_5(x_1+x_2+x_3;u^{-3},u^4)$,
 \item $P_5(x_1+x_2+x_3+x_4;iu^{-1},u)$,
 \item $P_{33}(x_1+x_2+x_3+x_4;u^{-3},u^2)$,
 \item $P_{33}(x_1+x_2+x_3+x_4+x_5;-u^{-2},u). $
\end{enumerate}
 Whilst the following polynomials are not highest weight Macdonald
 polynomials~:
 \begin{enumerate}
\item $P_4(x_1+x_2+x_3;u^{-1},u)$. Indeed,
 \[
 L_+^{q,t}(q,t)P_4(x_1+x_2+x_3;u^{-1},u)=3{(u+1)(1+u^2)(u^2+u+1)\over
 u^3}x_1x_2x_3.\]
 \item $P_5(x_1+x_2+x_3+x_4;u^{-1},u)$. Indeed ,
 {\footnotesize \[
 L_+^{q,t}P_5(x_1+x_2+x_3+x_4;u^{-1},u)=-4{(u+1)(u^2+1)(u^4+u^3+u^2+u+1)\over u^4}x_1x_2x_3x_4
 \]}
 \item $P_{33}(x_1+x_2+x_3+x_4+x_5;u^{-2},u)$. Indeed,
 {\footnotesize\[L_+^{q,t} P_{33}(x_1+x_2+x_3+x_4+x_5;u^{-2},u)=-2{(1-u^4)(1-u^6)(1-u^5)\over
 u^6(1-u)^2(1-u^2)}x_1x_2x_3x_4x_5.\]}
\end{enumerate}

 In conclusion, for each alphabet $\X$ of size $n$ and each rectangular
 partition $(k^r)$ with $n\geq 2r$, there is an explicit specialization of $(q,t)$ such that
$P_{k^r}(\X;q,t)$ belongs to the kernel of $L_+^{q,t}$.

\subsection{Staircase partitions}

 We examine here another family of highest weight Macdonald
 polynomials indexed by staircase partitions. By
 staircase partition, we mean a partition under the form 
$\lambda=[((\beta+1)s+r)^k,(\beta
 s+r)^l,\dots,(s+r)^l]$ where $\beta, s, r, k, l\in \N$ and $k\leq
 l$.

{\bf Theorem II.}

{\it 
Let $\beta, s, r, k, l\in \N$ with $k\leq l$. Consider the
partition $\lambda=[((\beta+1)s+r)^k,(\beta
 s+r)^l,\dots,(s+r)^l]$. The polynomial
 $P_\lambda(x_1+\dots+x_n;q,t)$ is a highest weight polynomial
 when
 \be
n={l+1\over s-1}r+l(\beta+1)+k
 \ee
 is an integer and
 \begin{equation}
\label{specqt}
  (t,q)=(u^{s-1\over g},u^{-{l+1\over g}}\omega_1)
 \end{equation}
 where $g=\gcd(l+1,s-1)$ and $\omega_1=\exp\left\{2i\pi(1+dg)\over
 s-1\right\}$ if $d$ denotes an integer such that $w_1^r=1$.\\
 Remark that the condition ${l+1\over s-1}r\in\N$ implies that
 ${s-1\over g}$ divides $r$ and the condition $\omega_1^r=1$
 implies $g$ divides $r$. Hence, in all the cases, $s-1$ divides
 $r$.}
\bigskip

{\bf Proof} As for the rectangular partition, the starting point
of our reasoning is the equality (\ref{L+qtP}):
{\footnotesize\begin{equation}\label{L+qtP2}
L_+^{q,t}.P_\lambda(\X;q,t)=\sum_i{1-t^{n-i}q^{\lambda_i}\over
 1-q}\prod_{j=i+1}^n{1-q^{\lambda_i-\lambda_j-1}t^{j-i+1}\over 1-q^{\lambda_i-\lambda_j}t^{j-i}}
 {1-q^{\lambda_i-\lambda_j}t^{j-i-1}\over
 1-q^{\lambda_i-\lambda_j-1}t^{j-i}}P_{\lambda_{(i)}}(\X;q,t).\end{equation}}
 Since $\lambda=[((\beta+1)s+r)^k,(\beta
 s+r)^l,\dots,(s+r)^l]$, there is only $\beta+1$ indices 
$i_v$ $(0\leq v\leq\beta)$ such that $\lambda_{(i_v)}$
 is a partition and hence has a non zero contribution in Eq. (\ref{L+qtP2}). These
 indices are characterized by $i_v:=k+lv$ and the corresponding
 node in the partition $\lambda$ is
\begin{equation}\label{node}(i_v,j_v)=(k+lv,(\beta+1-v)s+r).
\end{equation}
One verifies that $\lambda$ is weakly $(l,s,n)$-admissible. Hence,
Lemma 2 implies that $(t,q)=(u^{s-1\over
g},u^{-{l+1\over g}}\omega_1)$ is a pole  of neither
$P_\lambda(\X;q,t)$ nor $P_{\lambda_{(i_v)}}(\X;q,t)$ for
$v=0,\dots,\beta$. Hence, to prove the theorem it remains to show
that the coefficient $\varrho_v$ of $P_{\lambda_{(i_v)}}(\X;q,t)$
in Eq.(\ref{L+qtP2}) vanishes under the specialization (\ref{specqt}).\\
Let us examine first the denominator of $\varrho_v$. From Eq.
(\ref{L+qtP2}) this denominator is a product of polynomials under
the form $1-q^{\lambda_i-\lambda_j}t^{j-i}$,
$1-q^{\lambda_i-\lambda_j-1}t^{j-i+1}$,
$1-q^{\lambda_i-\lambda_j-1}t^{j-i}$ or $1-q$. Since $\lambda$ is
weakly admissible, Eq.(\ref{WFJMM1}) implies that the three first
possibilities do not vanish for the specialization (\ref{specqt})
whilst the fourth is straightforwardly not zero under this
specialization. Hence, it suffices to prove that the numerator
$\varsigma_v$ of $\varrho_v$ vanishes for each $v$.
One has to consider two cases.
First consider that $v=l$. After simplification, one obtains
\be
 \varsigma_l=(1-q^{r+s})(1-q^{s+r-1}t^{n-k-\beta l+1}).
\ee
But since $n-k-l\beta+1={l+1\over s-1}r+l-1$, it follows
\be
 q^{s+r-1}t^{n-k-\beta l+1}=q^rt^{{l+1\over s-1}r}=\omega_1^r=1,
\ee
from the hypothesis. Hence, $\varsigma_l=\varrho_l=0$.
Suppose now that $v<l$. One has
\be
 \varsigma_v=(1-t^{n-i_v}q^{\lambda_{i_v}})
\prod_{j=i_v+1}^n(1-q^{\lambda_{i_v}-\lambda_j-1}t^{j-i+1})(1-q^{\lambda_{i_v}-\lambda_j}t^{j-i-1}).
\ee
If we set $j=i_v+l$ then $\lambda_{i_v}-\lambda_j=s$ and the
factor $1-q^{\lambda_{i_v}-\lambda_j-1}t^{j-i_v+1}$ in $\varsigma$
vanishes under the specialization (\ref{specqt}). It follows that
$\varsigma_v=\varrho_v=0$. This implies our theorem. $\Box$

Let us give some examples. The following polynomials are highest
weight Macdonald polynomials:
\begin{enumerate}
\item $P_{53}(x_1+x_2+x_3+x_4+x_5;q=u^{-2},t=u)$ ($\beta=2$,
$s=2$, $k=0$, $r=1$, $l=1$).
 \item $P_{63}(x_1+x_2+x_3;q=-u^{-1},t=u)$ ($\beta=2$,
$s=3$, $k=0$, $r=0$, $l=1$).
 \item $P_{422}(x_1+x_2+x_3+x_4+x_5;q=u^{-3},t=u)$
 \item $P_{533}(x_1+\dots+x_8;q=u^{-3},t=u)$
 \item $P_{633}(x_1+\dots+x_5;q=u^{-3},t=u^2)$
 \end{enumerate}
 Note that the converse of Theorem (2) is false as shown
 by the counter-example~:
 \be
L_{+}^{q,t}P_{42}(x_1+\dots+x_2;q,t)=0
 \ee
 for $n\geq 2$ and $q=-1$.
 For the moment, the problem of the characterization of the
 highest weight Macdonald polynomials is still open.

\section{Highest weight Jack polynomials \label{HWJ}}

\subsection{Some necessary conditions}

   As it is shown in Ref.(\onlinecite{Lass}), the action of $L_+$ on the
   polynomials $P_\lambda^{(\alpha)}(\X)$ can be recovered from
   Eq. (\ref{L+qtP}) by setting $q=t^\alpha$ and sending $t$
   to $1$.
   {\footnotesize
 \begin{equation}\label{L+P}
  \begin{array}{rcl}L_+.P_\lambda^{(\alpha)}(\X)&=&\displaystyle\sum_i{n-i+\lambda_i\alpha\over
  \alpha}\times\\&&\times\displaystyle\prod_{j=i+1}^n{(\alpha(\lambda_i-\lambda_j-1)+j-i+1)(\alpha(\lambda_i-\lambda_j)+j-i-1)
  \over
  (\alpha(\lambda_i-\lambda_j)+j-i)(\alpha(\lambda_i-\lambda_j-1)+j-i)}P_{\lambda_{(i)}}^{(\alpha)}(\X)
  \end{array}
 \end{equation}}

If one asks the highest weight condition $L_+.J_\lambda^{(\alpha)}=0$
     in terms of $J_\lambda^{(\alpha)}$
    as in Ref. (\onlinecite{BH4}), then one has the necessary condition
    $n-l(\lambda)+1+\alpha(\lambda_{l(\lambda)}-1)=0$. However
for some specializations we may have $J_\lambda^{(\alpha)}=0$
    (for instance $J_{53}^{(-1)}(x_1+x_2+x_3+x_4)=0$).
To make sense, the property must be stated in terms of $P$:
 If $L_+.P_\lambda^{(\alpha)}(x_1+\dots+x_n)=0$ then
\begin{equation}
\label{lastpart}
n-l(\lambda)+1+\alpha(\lambda_{l(\lambda)}-1)=0.
\end{equation}
Indeed, we need that the coefficient of each
$P_{\lambda_{(i)}}^{(\alpha)}(\X)$ in Eq. (\ref{L+P}) vanishes. In
particular, the coefficient of
$P_{\lambda_{(l(\lambda))}}^{(\alpha)}(\X)$,
\[
\lambda_{l(\lambda)}(n-l(\lambda)+\alpha(\lambda_{l(\lambda)}-1)+1)\over
1+\alpha(\lambda_{l(\lambda)}-1)
\]
must equal $0$. Since $P_{\lambda_{(l(\lambda))}}^{(\alpha)}\neq
0$ occurs in Eq. (\ref{L+P}) ( $\lambda_{(l(\lambda))}$ being
always a partition and $P_{\lambda_{(l(\lambda))}}^{(\alpha)}$
being dominated by $m_{\lambda_{(l(\lambda))}}$), it follows that
one has necessarily
 $n-l(\lambda)+1+\alpha(\lambda_{l(\lambda)}-1)=0.$\\ \\
Set $\lambda=(r_m^{l_m},\dots,r_1^{l_1})$ with $r_m>\dots>r_1$.
Eq. (\ref{lastpart}) provides a necessary condition relying
$\alpha$ and $n$. It follows that if
$L_+.P_\lambda^{(\alpha)}(x_1+\dots+x_n)=0$,
 then $\alpha$ is a negative rational number and that the last part of $\lambda$ is strictly greater than $1$.
Other parts of $\lambda$ gives also further
information, which fixes the two values.
Suppose $n>l(\lambda)-1+l_1{r_1\over r_2-r_1}$ and
$\alpha\neq 0$.
 We examine the coefficient of
$P_{\lambda_{l(\lambda)-l_1}}^{(\alpha)}$ in Eq. (\ref{L+P}) after
simplification~:
\be
\alpha
(\alpha(r_2-r_1-1)+l_1+1)(r_2-r_1)(\alpha(r_2-1)+n+1-l(\lambda)+l_1)(\alpha
r_2+l_1)\over (\alpha(r_2-r_1)+l_1)(\alpha(r_2-r_1-1)+1)(\alpha
r_2+n+l_1-l(\lambda))(\alpha(r_2-1)+1+l_1).
\ee
If $L_+.P_\lambda^{(\alpha)}(\X)=0$, at least one of the five
factors $\alpha$, $(\alpha(r_2-r_1-1)+l_1+1)$, $(r_2-r_1)$,
$(\alpha(r_2-1)+n+1-l(\lambda)+l_1)$ or $(\alpha r_2+l_1)$
vanishes. From the hypothesis $\alpha>0$ and $r_2>r_1$. Hence, it
remains three factors: $(\alpha(r_2-r_1-1)+l_1+1)$,
$(\alpha(r_2-1)+n+1-l(\lambda)+l_1)$ and $(\alpha r_2+l_1)$.
Suppose $\alpha r_2+l_1=0$, Eq.(\ref{lastpart}) implies
\be
 n=l(\lambda)-1+l_1{r_1-1\over r_2}<l(\lambda)-1+l_1{r_1-1\over
 r_2-r_1}.
\ee
But this contradicts the hypothesis $n>l(\lambda)-1+l_1{r_1\over
r_2-r_1}$. In the same way, $\alpha(r_2-1)+n+1-l(\lambda)+l_1=0$
implies $n=l(\lambda)-1+l_1{r_1\over r_2-r_1}$ which also
contradicts the same hypothesis. It remains
$\alpha(r_2-r_1-1)+l_1+1$,
that is $\alpha={l_1+1\over 1-(r_2-r_1)}$.\\
Straightforwardly, this implies $1-(r_2-r_1)\neq 0$ or
equivalently $r_2>r_1+1$. Substituting $\alpha={l_1+1\over
1-(r_2-r_1)}$ in
$n-l(\lambda)+1+\alpha(\lambda_{l(\lambda)}-1)=0$, one obtains~:
\be
 n=l(\lambda)-1+{l_1+1\over r_2-r_1-1}(r_1-1).
 \ee
 Since $n$ is an integer, this implies that $r_2-r_1-1\over\gcd(r_2-r_1-1,l_1+1)$ divides $r_1-1$.

 In conclusion, under the condition
$n>l(\lambda)-1+l_1{r_1\over r_2-r_1}$ and $\alpha\neq 0$, if
$L_+.P_\lambda^{(\alpha)}(\X)=0$ then
\begin{equation}\label{r2}\alpha={l_1+1\over 1-(r_2-r_1)}.\end{equation}
 It follows that $r_2>r_1+1$
and $r_2-r_1-1\over\gcd(r_2-r_1-1,l_1+1)$ divides $r_1-1$. \\ In
others words, when $n$ is big enough, $\alpha\neq 0$ and for a
fixed
 partition $\lambda$ with at least two distinct parts, the
 polynomial $L_+.P_\lambda^{(\alpha)}(\X)$ vanishes for at most one
 value of $(n,\alpha)$.

\subsection{Rectangular  partitions}

  In this paragraph, we are interested in characterizing highest weight Jack
  polynomials indexed by rectangular partitions.
   One has, as a straightforward
  consequence of Eq.(\ref{lastpart}),
  \begin{equation}
\label{rectJack}
  L_+.P_{k^r}^{(\alpha)}(\X)=0\mbox{ implies }\alpha={r+1-n\over k-1}.
  \end{equation}
  Furthermore, if in addition  $n\geq 2r$ and $\gcd(n-r+1,k-1)=1$, 
a special case of Theorem (1) by  sending $u$ to $1$
  gives the equivalence of the two equalities.
Let us give some examples of such highest weight Jack
polynomials by computing their expansion over the basis $(\tilde
c_\lambda)_\lambda$. The simplest examples are provided by
partitions whose all the parts equals $2$. In this case, Eq.
(\ref{rectJack}) implies $\alpha=l(\lambda)-1-n$. The coefficients
seem easy to obtain and the first
computations suggest the general equality
\be
 P_{2^l}^{(1-l-n)}(x_1+\dots+x_n)=(-1)^{l}{n!^2\over
 (n-l)!^2}\sum_{\mu}(-2)^{l(\mu)}\tilde c_\mu,
\ee
summed over the partitions $\mu$ of $2l$ having only even parts.
Note that when $\gcd(n-r+1,k-1)\neq1$, $P_{k^r}^{(\alpha)}$ can
have a pole at $\alpha={r-1-n\over k+1}$, as shown by the example
\[\begin{array}{rcl}
P_{33}^{(\alpha)}(x_1+\dots+x_5)&=&m_{{3,3}}+3\,{\frac
{m_{{3,2,1}}}{2\,\alpha+1}}+6\,{\frac {m_{{3,1,1,1}}}{ \left
(2\,\alpha+1\right )\left (\alpha+1\right )}}+6\,{\frac
{m_{{2,2,2}}}{ \left (2\,\alpha+1\right )\left (\alpha+1\right
)}}\\&&+3\,{\frac {\left (3\,\alpha+5 \right )m_{{2,2,1,1}}}{\left
(2\,\alpha+1\right )\left (\alpha+1\right )^{2}}}+ 36\,{\frac
{m_{{2,1,1,1,1}}}{\left (2\,\alpha+1\right )\left (\alpha+1\right
)^{ 2}}},\end{array}
\]
$P_{33}^{(\alpha)}$ having a pole at $\alpha=-1$. Furthermore,
even correctly defined normalizations of $P_{33}^{(-1)}$ are not
annulled by $L_+$,
\be
 L_+.J_{33}^{(-1)}(x_1+\dots+x_5)=48m_{221}+288m_{2111}.
\ee
We conjecture the following property~:
 Suppose $n\geq r$, then
the assertions~:
  \be
P_{k^r}^{(\alpha)}(\X)\mbox{ is well defined and }L_+.P_{k^r}^{(\alpha)}(\X)=0
\ee
   and
\be
\alpha={r-1-n\over k-1},\ n\geq 2r\mbox{ and }n-r+1\mbox{ is not a divisor of }k-1
\ee
are equivalent.

\subsection{Staircase partitions}

    Let $\beta$, $s$, $r$, $k$ and $l$ be five integers such that
    $1\leq k\leq l$, $0<\beta,s$ and $\gcd(l+1,s-1)=1$. Let $\lambda=[((\beta+1)s+r)^k,(\beta s+r)^l,\dots,
    (s+r)^l]$ be a staircase partition and $\X$ an alphabet of
    size $n$ verifying $n>\beta l+k-1+l_1{s+r\over s}$. Let $\alpha\neq 0$ be a non zero complex number.
     From Theorem (2) and Eq. (\ref{lastpart}),
    the two following assertions are equivalent~:
    \begin{enumerate}
     \item \{$L_+.P_\lambda^{(\alpha)}(\X)=0$.\}
     \item \{$n={l+1\over s-1}r+l(\beta+1)+k  \quad{\rm and}\quad \alpha={1+l\over
     1-s}$.\}
    \end{enumerate}
Again, the implication $1\Rightarrow 2$ is a direct consequence of
Eq.(\ref{r2}). The implication $2\Rightarrow 1$ comes from a
special cases of Theorem (2) sending
$u$ to $1$.
The enumeration of the first cases 
allows us to propose the following conjecture~:
If $L_+.P_\lambda^{(\alpha)}(\X)=0$ then $\lambda$ is a staircase
partition.
If we assume the two previous conjectures, we can
propose a complete characterization of highest weight Jack
polynomials~:
 Let $\X$ be an alphabet of size $n$, $\alpha\neq 0$ and $\lambda$
 be a partition such that $l(\lambda)\leq n$. The polynomial $P_\lambda^{(\alpha)}(\X)$ is
 annulled by $L_+$ if and only if $\lambda$ is a staircase
 partition and one of the two following assertions is verified~:
 \begin{enumerate}
 \item \{If $\lambda=k^r$ is a rectangular partitions then $\alpha={r-1-n\over k-1}$, 
$n\geq 2r$ and $n-r+1$ is not a divisor of
 $k-1$.\}
 \item \{If $\lambda=[((\beta+1)s+r)^k,(\beta s+r)^l,\dots,
    (s+r)^l]$ is not rectangular, then $\gcd(l+1,s-1)=1$, $n={l+1\over s-1}r+l(\beta+1)+k $ and $\alpha={1+l\over
     1-s}$.\}
 \end{enumerate}
Remark that $\lambda=[((\beta+1)s+r)^k,(\beta s+r)^l,\dots,
    (s+r)^l]$ can be written in terms of occupation numbers as~:
\begin{equation}
\label{occ}
   \lambda\equiv
   [n_0,0^{s+r-1},l,0^{s-1},l,0^{s-1},\dots,l,0^{s-1},k,0^\infty].
\end{equation}
This includes the Jack polynomials indexed by partitions
$[n_0,0^{s+r-1},l,0^{s-1},l,0^{s-1},\dots,l,0^{s-1},l,0^\infty]$
of Ref.(\onlinecite{BH4}), but the last part may be different.
For instance, for $n=5$ and $\alpha=-3$, one has~:
\be
L_+.P^{(-3)}_{422}(x_1+x_2+x_3+x_4+x_5)=0.
\ee

\section{Conclusions}
\label{conclusion}

We have characterized highest weight Macdonald and Jack polynomials
for special partitions~: rectangular and staircase.
We have also formulated conjectures concerning a possible generalization.
To summarize, these conjectures  should be deduced from  that a 
necessary condition for a Macdonald polynomial to have a highest weight 
 is that its partition is a staircase, which is suggested by numerical evidences. 
The underlying mechanism seems to be related to the vanishing properties of 
staircase Macdonald polynomials under the specialization $q^at^b=1$. These 
vanishing properties could be translated in terms of factorizations under 
specializations of the variables $x_i$. For example, one has the identity~:
{\footnotesize
\be
\begin{array}{l}
P_{44}((1+t+t^2+t^3)x_1+(1+t+t^2)x_2;q=\omega t^{-2},t)=\\
(*)_{q,t}(x_1+\omega x_2)(x_1+\omega t^3x_1)(x_1-t^6 x_2)(x_2+\omega t^6x_1)
(x_1-t^9x_2)(x_2-t^9x_1)(x_2-t^{12}x_1)
\end{array}
\ee
}
where $(*)_{q,t}$ is a scalar depending only on $q$ and $t$, and 
$\omega=\exp\left\{2i\pi\over3\right\}$.
The link between highest weight and factorizations, generalizing the results of 
Ref. \onlinecite{BL},  is a promising study 
that will be explored in a future paper.
These wavefunctions may eventually prove useful for the construction
of candidate quasiparticle/quasihole states and their manipulation
by analytical or numerical~\cite{BR} means.

As a final remark, we note that similar results are known on non symmetric 
Jack polynomials (called ``singular'') which are in the kernel of the Dunkl 
operators~\cite{Dunkl1,Dunkl2}.
In this context, the study of singular non symmetric Macdonald 
polynomials seems to be relevant.

\begin{acknowledgments}

We wish to acknowledge Alain Lascoux for fruitful discussions about highest
weight symmetric functions in section \ref{HWSF}. 
We also acknowledge very useful interactions with A. Boussicault.
We thank also C.~F. Dunkl for pointing out the link between highest weight 
symmetric Jack polynomials and singular non symmetric Jack polynomials.

This paper is partially supported by ANR projects PhysComb, ANR-08-BLAN-
0243-04 and VolQuan, ANR-07-BLAN-0238.

\end{acknowledgments}


\appendix
\section{About some eigenvalue problems \label{Papen}}

The eigenvalue problem for a delta function interaction restricted
to the lowest landau level on symmetric functions admit only a handful
of explicit eigenstates~\cite{BPL,JK,SW,BP}. When the angular momentum
in the planar geometry is equal to the number of particles the lowest
energy ``yrast'' state is given by~:
  \be
 \Phi_{n}=\prod_{i=1}^n(x_1+\dots+x_n-nx_i).
  \ee
This quantity is a highest weight polynomial with a nice expression when written 
in terms of the functions $\tilde c_\lambda$~:
\be
\Phi_{n}=n^nn!\sum_\lambda (-1)^{l(\lambda)+n}\tilde c_\lambda
\ee
where the sum is over the partition $\lambda$ of $n$ having no
part equal to $1$. For example if $n=7$, one has
\be
\Phi_{7}=\prod_{i=1}^7(x_1+\dots+x_7-7x_i)=4150656720(\tilde
c_{{7}}-\tilde c_{{5,2}}-\tilde c_{{4,3}}+\tilde c_{{3,2,2}})
\ee
There also other exact states~\cite{BPL,JK,SW,BP} that are known~:
  \be
 \Phi^{n}_{L}=\sum_{i_1<\dots<i_L}\prod_{k=1}^L(x_{1}+\dots+x_{n}-nx_{i_k}).
  \ee
 Surprisingly,  this polynomial
 has the same expression (up to a multiplicative coefficient) in
 terms of $\tilde c_\lambda$ as $\Phi_{n}$:
\be
\Phi^{n}_{L}=
 n^LL!\left(n\atop L\right)\sum_\lambda
(-1)^{l(\lambda)+n}\tilde c_\lambda
\ee
(again the sum is over the partition $\lambda$ of $n$ having no
part equal to $1$). Note that the two polynomials are {\it not} equal
since they are evaluated on different alphabets. For example, we have~:
\be
\begin{array}{rcl}
\Phi^{9}_{7}&=&\displaystyle\sum_{1\leq i_1<i_2<\dots<i_7\leq
9}\prod_{k=1}^7(x_1+\dots+x_9-9x_{i_k})\\&=&867821895360 (\tilde
c_{{7}}-\tilde c_{{5,2}}-\tilde c_{{4,3}}+\tilde
c_{{3,2,2}})\end{array}
\ee

Note also that  the $\Phi^{n}_{L}$ are not the
symmetrized of the  $\Phi_{n}$ functions since their
expansions in terms of $\tilde c_\lambda$ are quite different.  For
example, the symmetrized of $\Phi_7$ on the alphabet $\{x_1,\dots,x_9\}$ is~:
\be
\begin{array}{l}
\displaystyle
\sum_{1\leq i_1<i_2<\dots<i_7\leq
9}\prod_{k=1}^7((x_{i_1}+\dots+x_{i_7})-7x_{i_k})=\\
486777211200\,\tilde c_{{7}}-462909867840\,\tilde
c_{{5,2}}-492922366272\,\tilde c_{{4,3}}+ 447162907968\,\tilde
c_{{3,2,2}}.
\end{array}
\ee


\section{Highest weight polynomials and the eigenvalues of the Read-Rezayi Hamiltonian \label{hk}}

The $k$-type Read-Rezayi
 state~\cite{Read96,Read99} is the exact zero energy ground state of smallest degree of~:
 \be
 {\cal
 H}^{(k)}:=\sum_{i_1<\dots<i_{k+1}}
\delta^{(2)}(x_{i_1}-x_{i_2})
\dots
\delta^{(2)}(x_{i_k}-x_{i_{k+1}}).
 \ee
In the lowest Landau level we are only interested by
the description of the spectral properties of the operator~:
\be
{\bf h}_k.f(x_1,\dots,x_n)=\sum_{i_1<\dots<i_k}
f(x_{1},\dots,x_{i_1-1},X,x_{i_1+1},\dots,,x_{i_{k+1}-1},X,x_{i_{k+1}+1},\dots,x_n),
\ee
where $X$ denotes $X={x_{i_1}+\dots+x_{i_{k+1}}\over k+1}$, acting on
symmetric functions.
The computation of the eigenspaces of the operator ${\bf h}_k$
is highly non trivial since its characteristic polynomial
generally does not factorize in the field of rational numbers. 
One eigenfunction can be easily shown for any $k$~:
the sum of the variables $c_1=x_1+\dots+x_n$. More precisely, a
straightforward computation gives~:
\be
{\bf h}_k.c_1=\left(n\atop k+1\right)c_1.
\ee
Furthermore, ${\bf h}_k$ commutes with the multiplication by
$c_1$:
\be
 {\bf h}_kc_1f(x_1+\dots+x_n)=c_1{\bf h}_kf(x_1+\dots+x_n).
\ee
The equality $\tilde c_1=c_1$ combined to the fact that ${\bf h}_k$
is diagonalisable implies that it suffices to understand the
eigenspaces of the restriction of ${\bf h}_k$ to the space
generated by the $\tilde c_\lambda$ where $\lambda$ is a partition
without $1$, that is the
algebra of highest weight symmetric polynomials.
Note that in the special case $k=1$, the polynomials $\Phi^{n}_{L}$
are eigenfunctions of ${\bf h}_1$ with eigenvalues
$\frac12L(L-{n+2\over2})$.


In fact it is enough to find the highest weight
eigenfunctions of ${\bf h}_k$. Let us illustrate this principle
with the simplest example $n=3$ and $k=1$. This is a particularly
simple case, since the characteristic polynomial factorizes. One
has to find as many eigenfunctions as the numbers of partitions
with parts only equal to $2$ or $3$ which is given by the
generating function~:
\be\begin{array}{rcl}
\frac1{(1-t^2)(1-t^3)}&=&1+t^2+t^3+t^4+t^5+2t^6+t^7+2t^8+2t^9+2t^{10}+2t^{11}+\\
&&3t^{12}+2t^{13}+3t^{14}+3t^{15}+3t^{16}+3t^{17}+4t^{18}+3t^{19}+\cdots.\end{array}
\ee
The square of the Vandermonde determinant
$(x_1-x_2)^2(x_1-x_3)^2(x_2-x_3)^2$ belonging to the kernel, it
remains to compute as many functions as the numbers described by
the generating series
\be
\frac1{(1-t^2)(1-t^3)}-\frac{t^6}{(1-t^2)(1-t^3)}=
1+\frac{t^2}{(1-t)}=1+{t}^{2}+{t}^{3}+{t}^{4}+{t}^{5}+{t}^{6}+{t}^{7}+{t}^{8}+\dots,
\ee
\emph{i.e.} only one by degree. We conjecture that the following
polynomials are eigenfunctions of ${\bf h}_1$:
{\footnotesize
\be
 \Psi^{(2)}_L:=\tilde c_{3^p2^\ell}+6(p+\ell)\tilde
c_{3^{p-2}2^{\ell+3}}+36(p+\ell)(p+\ell+1)\tilde
c_{3^{p-4}2^{\ell+6}}+...+6^{q}(p+\ell)...(p+\ell+q)\tilde c_{3^\epsilon
2^{\ell+3q}},
\ee
}
where $p$ is the maximal integer such that $L=3p+2\ell$ with $0\leq
\ell$ and $p=2q+\epsilon$ with $\epsilon=0$ or $1$. 

Some examples are~:
{\footnotesize
\be L=21:\, \tilde c_{3333333}+42\tilde c_{33333222}+2016\tilde c_{333222222}+108864\tilde c_{3222222222},\ee
\be L=22:\, \tilde c_{33333322}+48\tilde c_{333322222}+2592\tilde c_{3322222222}+155520\tilde c_{22222222222},\ee
\be L=23:\,
\tilde c_{33333332}+48\tilde c_{333332222}+2592\tilde
c_{3332222222}+155520\tilde c_{32222222222}.\ee
}
Unfortunately, the general case is not so simple. But, we hope
that the subproblem of the description of the kernels  can be
solved more easily by means of a similar reasoning. The difficulty
consists in finding a ``good'' family of symmetric functions such
that a basis of the kernel can be ``nicely'' described. 
Numerical evidences suggest that the only highest weight Jack polynomials belonging to the kernel of ${\bf h}_k$
are rectangular. Furthermore, the first computations suggest that these polynomials play an 
important role in the description of the kernel. For example, when  $n=4$, the restriction 
of the kernel of ${\bf h}_2$ to the space of highest weight polynomials is generated by two 
algebraically independent polynomials $P^{(-3)}_{22}$ and $P^{({-\frac32})}_{33}$. 
But the construction is not understood in the general case, for example when $n=5$ the 
kernel of ${\bf h}_4$ (restricted to highest weight polynomials) is generated by 
$P_{22}^{(-3)}$, ${\cal S}P_3^{(-\frac32)}(x_1+x_2+x_3)P_2^{(-2)}(x_4+x_5)$ and 
${\cal S} P_{33}^{(-\frac32)}(x_1+x_2+x_3+x_4)$, where ${\cal S}$ means symmetrization 
\emph{w.r.t.} the alphabet $x_1+x_2+x_3+x_4+x_5$.


\end{document}